\documentclass[aapm,mph,preprint,graphicx]{revtex4-1}
\draft
\usepackage[mathlines]{lineno}

\usepackage{ragged2e}
\usepackage{graphicx}
\usepackage{lipsum}
\usepackage{amsmath}
\usepackage{indentfirst}
\usepackage{multirow}

\usepackage{xcolor}
\begin{document}

\title{Examining the Influence of Digital Phantom Models in Virtual Imaging Trials for Tomographic Breast Imaging} 

\justify
\author{Amar Kavuri$^1$}

\author{Mini Das$^{1,2}$}

\email[Email: ]{mdas@uh.edu}
\thanks{Author to whom correspondence should be addressed.}
\affiliation{$^1$Department of Biomedical Engineering, University of Houston, Houston,TX-77204,USA}

\affiliation{$^2$Department of Physics, University of Houston, Houston,TX-77204,USA}

\date{\today}
\justify
\begin{abstract}

\textbf{Purpose:} 
Digital phantoms are one of the key components of virtual imaging trials (VITs) that aims to assess and optimize new medical imaging systems and algorithms. However, these phantoms vary in their voxel resolution, appearance and structural details. This study aims to examine whether and how variations between digital phantoms influence system optimization with digital breast tomosynthesis (DBT) as a chosen modality.

\textbf{Methods:}
We selected widely used and open access digital breast phantoms generated with different methods. For each phantom type, we created an ensemble of DBT images to test acquisition strategies. Human observer localization ROC (LROC) was used to assess observer performance studies for each case. Noise power spectrum (NPS) was estimated to compare the phantom structural components. Further, we computed several gaze metrics to quantify the gaze pattern when viewing images generated from different phantom types.

\textbf{Results:} 
Our LROC results show that the arc samplings for peak performance were approximately $2.5^\circ$ and $6^\circ$ in Bakic and XCAT breast phantoms respectively for 3-mm lesion detection task and 
indicate that system optimization outcomes from VITs can vary with phantom types and structural frequency components. Additionally, a significant correlation (p<0.01) between gaze metrics and diagnostic performance suggests that gaze analysis can be used to understand and evaluate task difficulty in VITs.

\textbf{Conclusion:}
Our results point to the critical need to evaluate realism in digital phantoms as well as ensuring sufficient structural variations at spatial frequencies relevant to the signal size for an intended task. In addition, standardizing phantom generation and validation tools might aid in lower discrepancies among independently conducted VITs for system or algorithmic optimizations.
\end{abstract}


\maketitle

\section{INTRODUCTION}\label{intro}
	Rapid advances in medical imaging technologies and methods make it impossible to evaluate and optimize several emerging and competing systems and algorithms using clinical imaging trials. These trials can take extensive resources and long duration. Virtual imaging trials (VITs) are alternative approach to assess the potential of an imaging system, software or specific system/software combinations.\cite{bakic2013mo,badano2017silico,das2011comparison,das2015examining,barufaldi2018openvct,sharma2019silico,abadi2020virtual,barufaldi2021virtual} VITs are based on in-silico methods, where digital phantoms replace patients, software platform mimics imaging process, and virtual interpreters represent the human readers. 

As VITs are becoming more accurate, realistic digital phantom development has drawn much attention from the medical imaging community.  Our group recently examined the contribution of anatomical and quantum noise in signal detection and performance for human observers in such imaging trials \cite{kavuri2020relative,kavuri2018interaction}. Using VITs, we have also examined in the past, the development of novel visual search observer models to match humans \cite{gifford2016visual,lau2013towards,jiang2017analyzing}, system and algorithm optimization questions \cite{gifford2008optimizing,das2015examining,das2011comparison,das2008evaluation,das2009evaluation,gifford2018assessment}, understanding image texture features as relevant to human observer performance \cite{nisbett2020correlation, nisbett2018investigating}, and radiomics variability \cite{andrade2022sources}. As in other group's work, all of these studies were conducted with one breast phantom type.

Here we will evaluate what implications a VIT study outcome may have by changing the phantom type.  We once again take DBT as the modality of choice to examine this critical question and use two widely accepted types of breast phantoms - both considered anthropomorphic. For the last few years, many computational breast phantoms have been developed using different methods.\cite{glick2018advances} Some commonly used breast phantoms include power-law based phantoms\cite{sechopoulos2009optimization,gang2010anatomical,lau2012statistically,baneva2017evaluation}, anthropomorphic phantoms\cite{bakic2002mammogram,bliznakova2010evaluation, chen2011anthropomorphic,mahr2011three,elangovan2017design,carton2014virtual}, modified patient tomographic data\cite{hsu2013generation,erickson2016population,sturgeon2016eigenbreasts,chen2017high,sarno2021dataset}, and mastectomy specimens\cite{chawla2009optimized,michael2013generation,o2010development}.
These phantoms vary in their resolution, model, structures, and details. Power spectrum analysis has been used as a method to assess phantom structures and realism. \cite{bliznakova2010evaluation, chen2011anthropomorphic,mahr2011three,elangovan2017design,michael2013generation,bakic2011development,cockmartin2013comparative}
 Cockmartin et al \cite{cockmartin2013comparative} observed in 2013 that none of the evaluated phantoms matched with the patient data in terms of power-spectrum parameters. Few studies extended the realism assessment with trained human's (physicists and radiologists) rating. \cite{bliznakova2010evaluation,elangovan2017design,hsu2013generation,sturgeon2016eigenbreasts}
 Badano\cite{badano2017much}, however, argued that realism is subjective and simulating relevant properties for the task is sufficient. But it is unclear which properties are relevant and what level of realism is sufficient for the task of tomographic imaging system/algorithmic optimizations.

Past work by various groups have used different  phantom types for evaluating DBT system configurations\cite{sechopoulos2009optimization,chawla2009optimized,reiser2010task,zeng2017optimization}. Because of the differences in phantoms, configurations, and interpreters among the studies, there is little agreement on the optimal configuration. Specifically, the contribution of phantom differences in these inconsistencies is not well understood. In preliminary analysis, Park et al. \cite{park2010statistical} and Zhao et al \cite{zhao2014noise} observed that phantom differences (uniform vs structured) influence tomographic system optimal configurations. However, differences resulting from various available structured phantoms (when they are each considered realistic by the research community) has never been explored. 

We explore this critical question and examine whether and how phantom structure variability would influence the study outcome of a VIT such as for DBT acquisition parameters. In addition, we also investigated the influence of task difficulty due to change in the phantom structures using analysis of observer gaze pattern and interpretation process when viewing images generated with different phantoms in the VIT studies.  We present results from our study examining the influence of breast phantom types in VIT for DBT when all other aspects remain the same. To the best of our knowledge, ours is the first study to evaluate this critical aspect.

To accomplish this, we selected two widely used and open access breast phantom types which were generated using different procedures. Further, human observers analyzed simulated in-plane DBT images of the selected phantoms for similar configurations. Our results show a comparison of predicted system optimizations between VITs using these two phantom types.  

\section{MATERIALS AND METHODS}
The study methods are summarized below. First, the selected phantoms are listed. Second, our simulation methodology is described to generate both abnormal and normal cases of DBT images. Next, the experimental method to estimate human observer performance and gaze pattern is discussed. Finally, the procedure to characterize phantom structures using a power spectrum is described.

\subsection{Phantom Selection}
We selected two types of digital breast phantoms for this study: three-dimensional anthropomorphic breast models generated by Bakic et al.  at the university of Pennsylvania \cite{bakic2002mammogram} (hereafter referred to as Bakic phantom) and XCAT breast phantoms generated at Duke university using compressed volumes of patient breast CT data \cite{erickson2016population} (hereafter referred to as XCAT breast phantoms). Bakic phantoms are based on mathematical models which define the breast structural variability. These phantoms can be manipulated easily through the model configurations to simulate changes in anatomy. XCAT breast phantoms are based on patient breast CT images. These phantoms may appear more realistic, but changing their anatomical variations or resolution is difficult \cite{abadi2020virtual}. For our analysis, six digital phantoms of 5-cm thickness for each type were selected. The phantoms of each type were categorized based on volumetric glandular fractions (VGFs) with 25\% density and other three with 50\% density.
Bakic and XCAT breast phantoms have a voxel resolution of $(0.2 mm)^3$ and $(0.25 mm)^3$ respectively. The selected case numbers of XCAT phantoms of approximately 25\% density are CTA1608, CTA0357, and CTA1326 and of approximately 50\% density are CTB6013, CTA1284, and CTA1285.

\begin{figure}[ht!]
\centering
\includegraphics[width=0.4\textwidth]{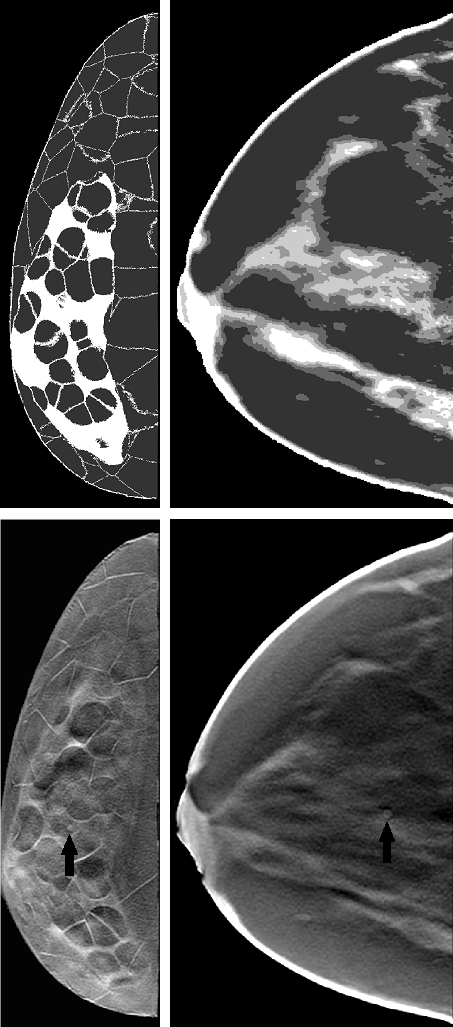} 
\caption{\label{fig:phantoms}A sample slice of 25\% VGF Bakic (top left) and XCAT breast (top right) phantoms and the corresponding 1-mm DBT slices with 3-mm spherical lesion on the bottom.}
\end{figure}

Figure \ref{fig:phantoms} shows sample slices of both types of phantoms on the top. The transitions in XCAT breast phantoms from 100\% glandular to adipose tissue have intermediately dense (25\%,50\%, and 75\%) voxels while these transitions are sharp (100\% dense to adipose) in Bakic phantoms. The XCAT breast phantoms lack small structures such as cooper`s ligaments. 
Erickson et al. \cite{erickson2016population} noted that the "lack of very fine-detail structures like cooper`s ligaments can negatively impact the realism of digital phantoms". References \cite{sturgeon2016eigenbreasts,chen2017high,rajagopal2018evaluation} tried different methods to improve XCAT breast phantoms, but these phantoms were not released publicly. 

\subsection{Image Generation} \label{images}
The DBT images used in this study were generated using a simulation platform based on serial cascade model\cite{siewerdsen1997empirical,vedula2003computer}.
The simulation platform modeled generic DBT systems with both source and detector rotating geometry, which was detailed in our previous work \cite{ kavuri2020relative,vedula2003computer,das2011penalized} and is described briefly here. 
The x-ray spectrum modeled a 30-kVp molybdenum anode source with 0.7-mm thick Al filter and the x-ray fluence scaled to provide a 1.5 mGy mean glandular dose (MGD) to a 5-cm-thick compressed breast. This total dose was evenly distributed at each projection angle and the x-ray fluence per projection was determined based on the breast dosimetry data (Dgn coefficients) generated by a Monte Carlo simulator\cite{boone1999glandular}. Focal-spot blurring with a 0.1-mm focal-spot size was modeled using a Gaussian modulation transfer function. The detector was modeled as a 0.1-mm thick CsI based a:Si flat panel detector with a 0.1-mm pixel size. The scintillator blurring was modeled using an empirically measured pre-sampling MTF. Quantum noise was modeled by a Poisson distribution for each keV (at the absorption of x-rays with in the scintillator), while additive electronic noise followed a Gaussian process with a standard deviation of 2200 electrons. Scatter was not modeled in this simulation.

The lesion targets were homogeneous spherical masses with a 3-mm diameter. 
While this is smaller than the average lesion sizes detected by the current DBT systems, VITs are also aimed at evaluating future imaging system designs.
This signal size of 3-mm was also chosen to achieve sufficient task-challenge without requiring to artificially alter attenuation values of the signal and to the capture the human observer performance trends more accurately for a range of breast densities.
An abnormal case was generated by substituting the lesion into the randomly selected location in the glandular region prior to the projection imaging. 
The lesion contrast or local glandularity were not matched between phantom types while selecting lesion locations as absolute performance was not relevant for this study.
The simulated mass was assumed to have an energy-dependent attenuation coefficient for invasive ductal carcinoma as reported by Johns and Yaffe.\cite{johns1987x}
Eight abnormal cases and one normal case were formed for each phantom.
The projections were acquired using Siddon's ray tracing method \cite{siddon1985fast} to model x-ray transmission through the breast. Two different sets of projections were generated with different phantom types. In order to filter the random noise, an adaptive Wiener filter based denoising algorithm was applied on each projection.\cite{lim1990two,vieira2013effect} The denoising algorithm was shown to reduce the noise effectively in our prior work \cite{kavuri2020relative}. Each data set was acquired over an angular span of $60^\circ$ with projection number $P\in$\{3, 7, 11, 15, 19, 21, 25, 31, 35, 41, 45\} by keeping the total dose steady at 1.5 mGy for each DBT acquisition. Feldkamp filtered back-projection (FBP) algorithm\cite{feldkamp1984practical} was used for image reconstruction. A three-dimensional Butterworth filter with a cutoff of frequency 0.25 cycles/pixel was applied on reconstructed volumes. In-plane DBT images of 1-mm thickness were produced by applying a boxcar averaging. 
Eight lesion-present (abnormal) and eight lesion-absent (normal) images were created for each phantom for the human observer studies as described in our earlier work\cite{das2015examining}. 
A set of 96 images were produced from the six phantoms of each phantom type for a given projection number.
Figure \ref{fig:phantoms} shows sample 1-mm abnormal DBT slices of both types of phantoms on the bottom.

\subsection{Human Observer Study}
Three non-radiologists took part in the Localization ROC (LROC) experiments. An LROC study entails both detection and localization of a 3-mm spherical lesion. 
The task in our study is search and localization/detection of spherical mass in simulated in-plane DBT images thus justifying the non-radiologists as observers.
Physicists and engineers who participated as observers had the same level of experience in reading simulated images. The human observer experimental method was same as that described in our prior work \cite{das2015examining,das2011comparison,lau2013towards,gifford2016visual} and summarized briefly here. In this study, in-plane DBT slices of the Bakic phantom set correspond to five acquisition protocols with projections of 7, 11, 19, 25 and 35 and XCAT breast phantom set correspond to six acquisition protocols with projections of 3, 7, 11, 19, 25 and 35 were evaluated. 
The 96 images per set (48 pairs of abnormal/normal images) were divided into 72 test images (36 pairs) and 24 training images (12 pairs). All sets included an initial training session followed by a test session. Each observer thus read 10 image sets. Observers were asked to select the lesion location, and a four-point ordinal scale was used to collect the confidence rating. Localizations were considered as correct when the observer selected the location within a 2-mm radius of lesion center (radius of spherical lesion + 0.5mm additional radius for human selection error). The observers performance was quantified with area under LROC curves (AUC). The estimate of AUC for a given observer and protocol was obtained with a Wilcoxon-based non-parametric ranking method. In order to assess the consistency of the AUC values between the observers, the intraclass correlation coefficient (ICC) was calculated using Python (v3.7.6) software (www.python.org).

\subsection{Eye Gaze Collection}
\begin{figure}[h!]
     \centering
	\includegraphics[width=0.45\textwidth]{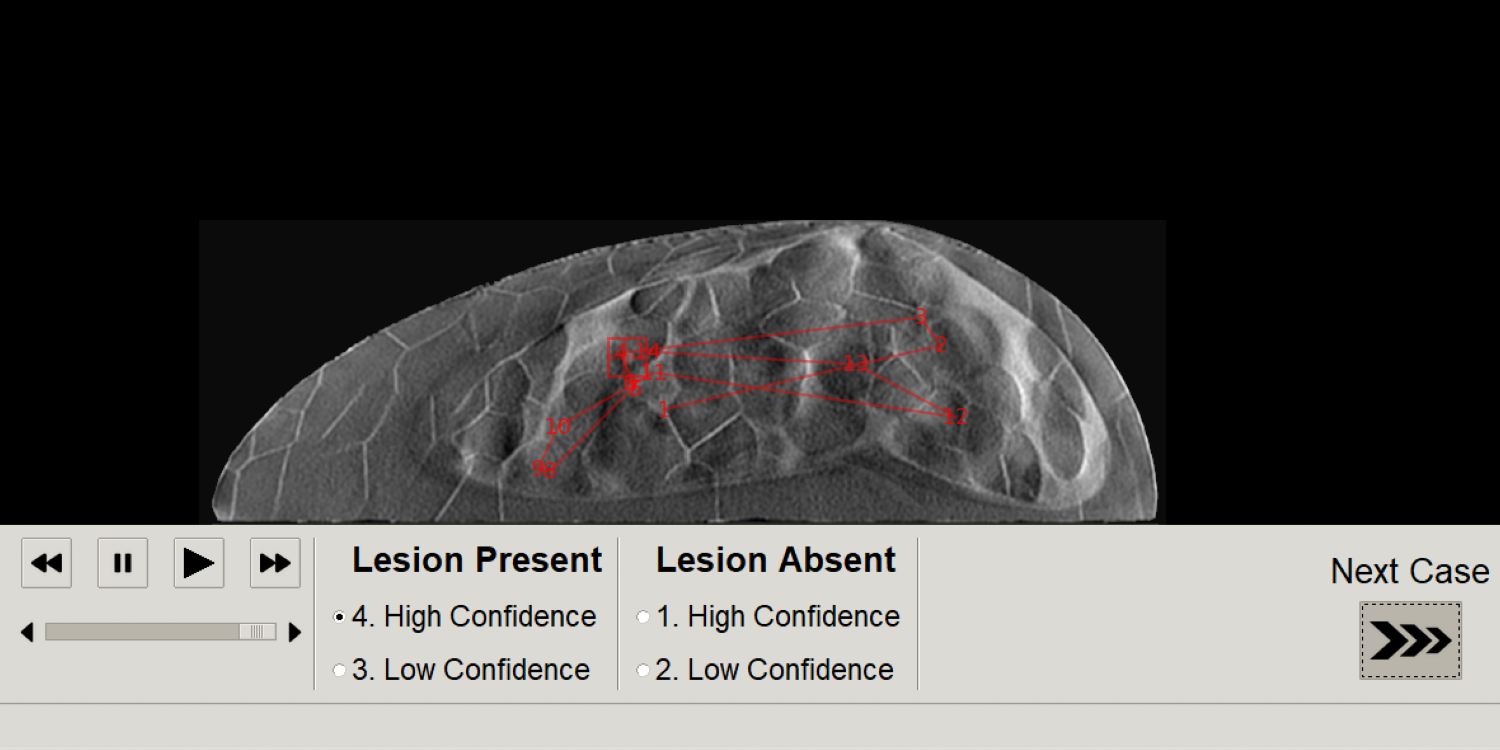}\vspace{\fill}\\
	\includegraphics[width=0.45\textwidth]{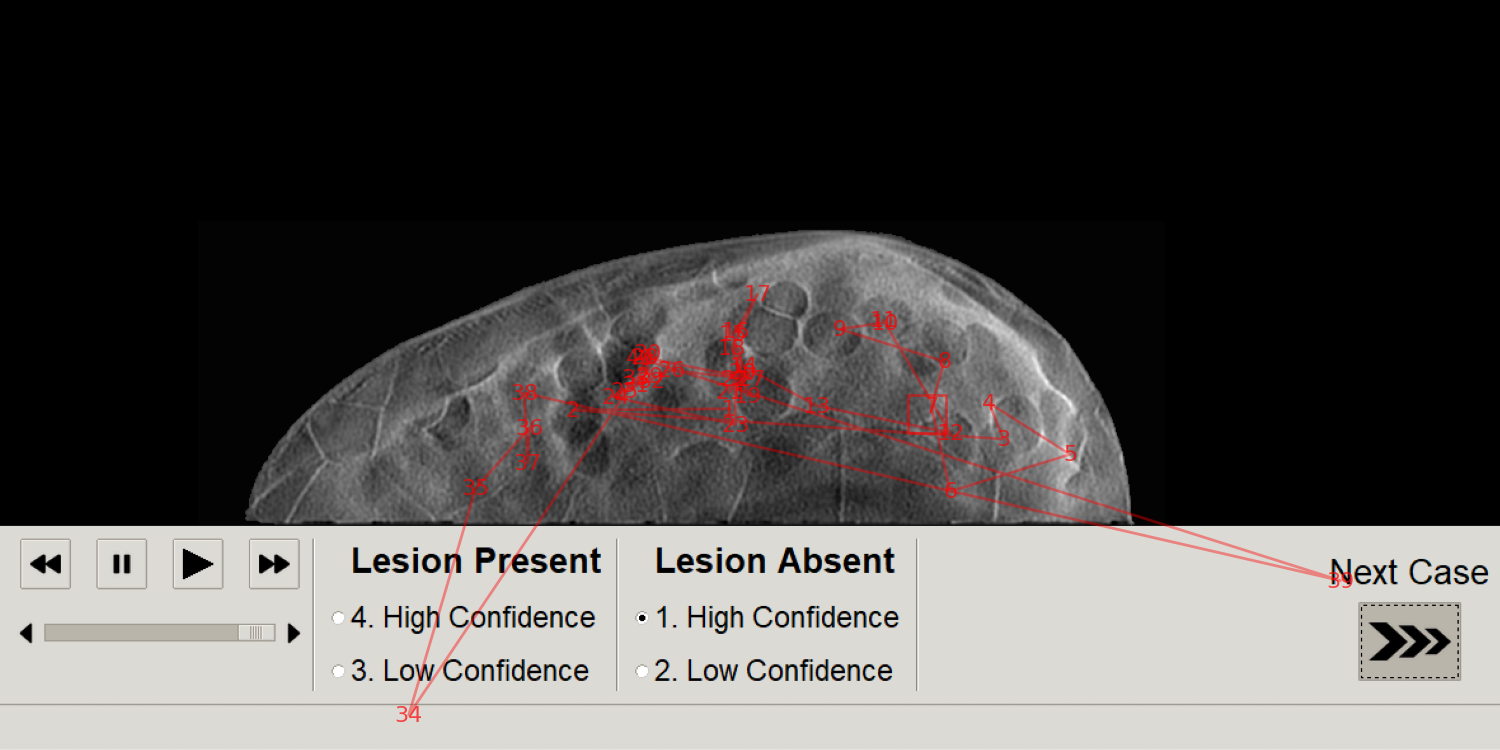}
        \caption{\label{fig:scan_ex}Example gaze pattern with fixations and saccades on DBT slices of 25\% dense (top) and 50\% dense (bottom) phantoms. }
\end{figure}

Numerous studies have demonstrated that gaze metrics correlate with diagnostic performance and can reveal human observers interpretation process\cite{brunye2019review,kundel1978visual,kundel2008using,voisin2013investigating}. In order to evaluate the differences in the interpretation process due to the change of phantom type, we collected gaze data using a screen based eye-tracking system (Tobii pro X3-120 with EPU ). An additional three more non-radiologists participated for the eye tracking study. The eye gaze data was collected for a single acquisition protocol of 35 projections for both type of phantoms. Both presentation of stimuli and eye-tracking were controlled by a Lenovo thinkpad p52s laptop using an in-house built Python software. A calibration and validation procedures were performed before starting a session. The images were displayed on a standard Dell 23.8 inch LCD monitor with a resolution of 1920 $\times$ 1080. Once the software collects and stores the gaze logs, post processing was applied to estimate fixation locations, fixation durations, and saccade durations using an I-VT filter based on Tobii pro white paper\cite{olsen2012tobii}. Figure \ref{fig:scan_ex} shows a sample gaze pattern of an observer for the task of searching and locating a 3-mm lesion. Each vertex indicates the fixation center with numbers indicating fixation order and  lines indicating the saccade paths and lengths.

We computed five gaze metrics to characterize the gaze pattern. Namely, we found total time spent on each image, total number of fixations made on each image, time taken to first fixate on the lesion region (first hit time), number of fixations on lesion region, and accumulated lesion dwell time. These gaze metrics were estimated using MATLAB2018a (The MathWorks, Inc.) with in-house built scripts. 
Spearman rank correlation was estimated between the average values of each gaze metric and the corresponding AUC values of the six observers.

\subsection{Noise Power Spectrum (NPS)}
Noise power spectrum (NPS) based analysis has been used in literature to quantify the similarity between phantom structures and clinical data.\cite{cockmartin2013comparative,chen2011anthropomorphic}. 
Our NPS estimation methods are briefly summarized here and can be found in detail elsewhere\cite{kavuri2020relative}. We selected multiple regions of interest (ROIs) of size 2.43cm$\times$2.43cm from each lesion-absent DBT slice from the breast region. The mean of each ROI was subtracted from the corresponding ROI and a Hann tapering window was applied to each ROI to reduce the edge artifacts. The 2D NPS was calculated by ensemble averaging the square of the magnitude of the discrete Fourier transform of each tapered ROI , and radial averaging of the 2D NPS was performed, resulting in a 1D NPS. A linear regression fit (from lower frequency ranges between 0.1 and 0.15 $mm^{-1}$ to higher frequency ranges between 0.4 and 0.7 $mm^{-1}$) was estimated to the natural logarithm of the 1D NPS that maximizes the coefficient of determination ($R^2$). The NPS parameter $\beta$ is estimated as the slope of linear fit.

\section{RESULTS}\label{results}
 \subsection{Human Observer LROC Results}
\begin{figure}[ht!]
\centering
\includegraphics[width=0.95\textwidth]{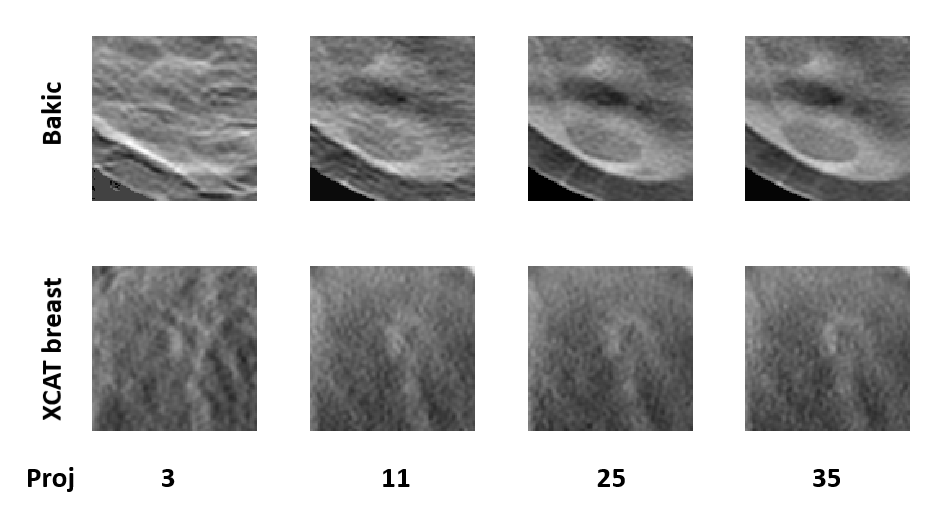} 
\caption{\label{fig:sampleROIs}A sample lesion present regions of DBT slices of both phantoms acquired with different number of projections.  }
\end{figure}

\begin{figure}[h!]
     \centering
\includegraphics[width=0.45\textwidth]{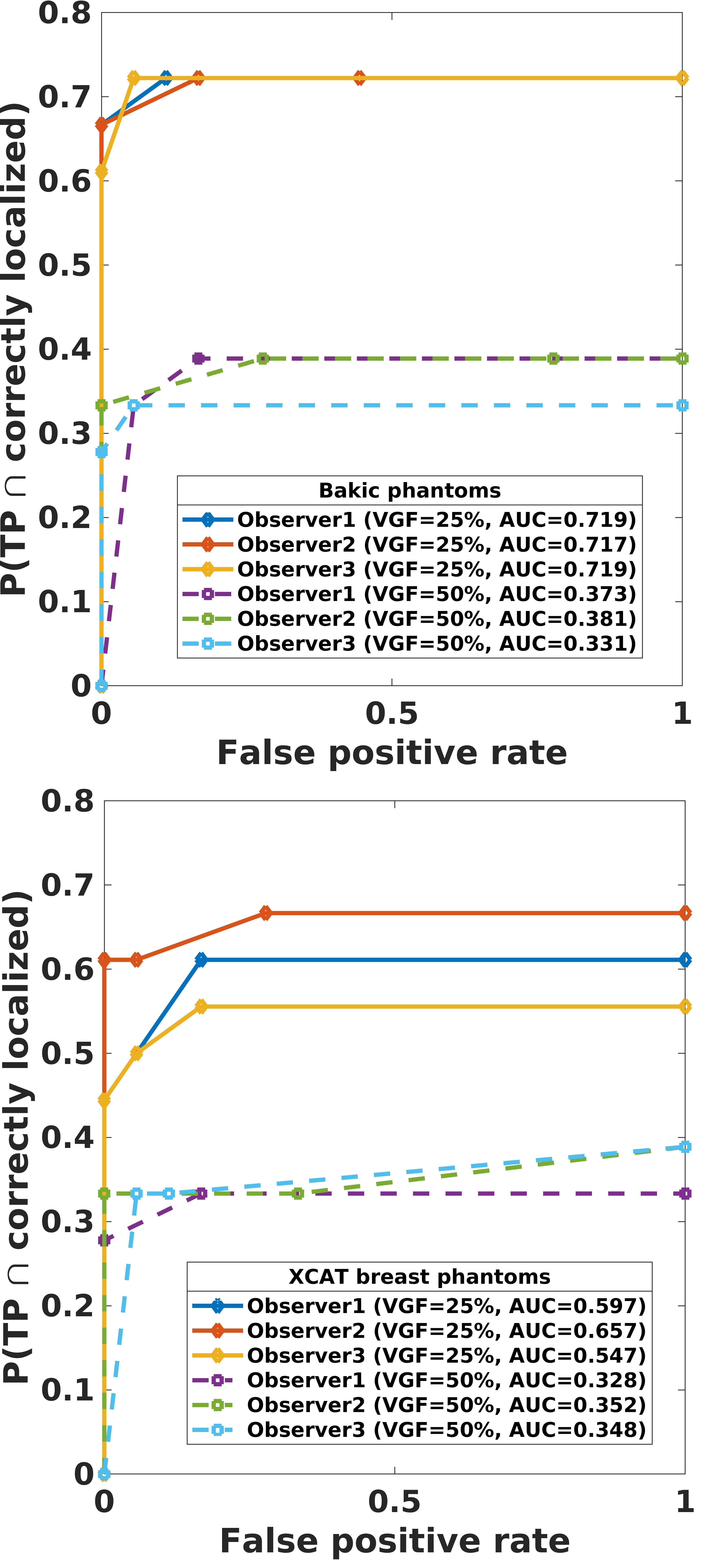} 
\caption{Human observer performance plotted as localization ROC (LROC) curves for a sample acquisition of 35 projections over $60^\circ$  in 3-mm mass detection study in Bakic phantom (top) and XCAT breast phantom (bottom) backgrounds.  }
\label{fig:lroc_fig}
\end{figure}

\begin{figure}[h!]
     \centering
\includegraphics[width=0.45\textwidth]{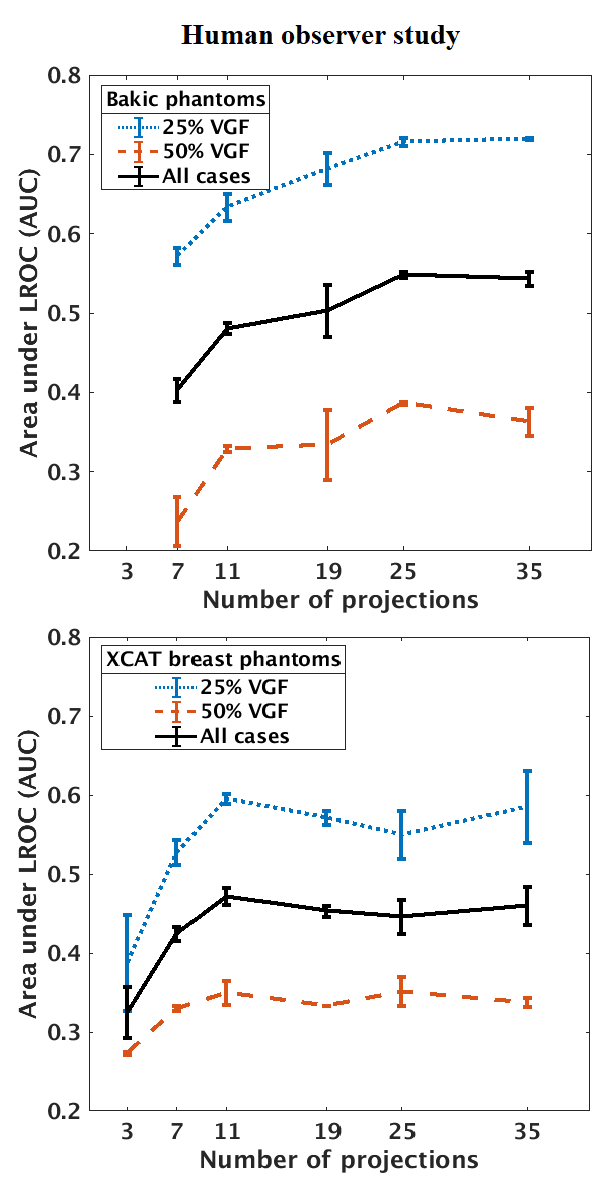} 
\caption{Human observer performance plotted as area under LROC (AUC) against number of projections in 3-mm mass detection study in Bakic phantom (top) and XCAT breast phantom (bottom) backgrounds. The results suggest the optimal configurations does not change with the task difficulty but changes with the background structure type. }
\label{fig:auc_fig}
\end{figure}

In our studies, each human observer was tasked with search and localization of a 3-mm lesion within the displayed DBT image slice (similar to the ones shown in fig \ref{fig:phantoms}). 
Figure \ref{fig:sampleROIs} shows sample regions of DBT slices with lesion at the center of region to illustrate the changes in the visibility of the signal for varying number of projections in both types of phantom backgrounds. The signal is better visible in Bakic phantom background for the acquisition configuration of 35 projections, whereas the signal is better visible in XCAT breast background for the acquisitions of 11 to 35 projections. 
Figure \ref{fig:lroc_fig} presents LROC plots for the three observers, for a sample acquisition of 35 projections over $60^\circ$ arc span, in both phantom backgrounds.
The y-axis of the LROC represents the joint probability of correctly localizing a lesion in a case reported as positive. Therefore, the curve reaches up to the percentage of cases with correct lesion localization. The LROC AUC values above zero are considered as better than guessing as the likelihood of guessing the lesion's location is zero.
Figure \ref{fig:auc_fig} shows the average performance of the three observers in both Bakic and XCAT breast phantom images. Error bar lengths indicate twice the standard error of the three observers' AUC values. We observed a greater improvement in performance up to 11 projections in XCAT breast phantom backgrounds and a steady performance thereafter. This corresponds to an arc sampling of approximately $6^\circ$ between adjacent projections for peak performance. In Bakic phantom backgrounds, observers' performance improved up to 25 projections and required finer arc sampling of approximately $2.5^\circ$ to achieve peak performance. 
We also plotted the detection performance separately for 25\% dense and 50\% dense slices, where 25\% dense indicates easy level of task difficulty while 50\% dense indicates higher level of task difficulty.
 Regardless of phantom type, both levels of task difficulty show similar trends and suggest that optimization may not change with the task difficulty, which is in accordance with earlier observations by Zeng et al. \cite{zeng2017optimization} and Mackenzie et al.\cite{mackenzie2021effect}.
 We also noticed in figures \ref{fig:lroc_fig} and \ref{fig:auc_fig} that observers had overall slightly lower performance in XCAT breast backgrounds than in Bakic backgrounds in particular for  25\% dense images. This difference can not be attributed to phantom structures necessarily as the local densities in the region where lesions are inserted could also influence these LROC AUC. The magnitude of AUC values are less relevant in this particular study as only relative change in AUC values with changing system parameters (such a number of projection in this study) were used to deduce the final conclusions.
The inter-observer agreement, quantified using the ICC, ranged from 0.92 to 0.95 with the average AUC values, suggesting a strong agreement between the observers. 

\subsection{Power Spectrum Analysis}
\begin{figure}[h!]
     \centering
\includegraphics[width=0.45\textwidth]{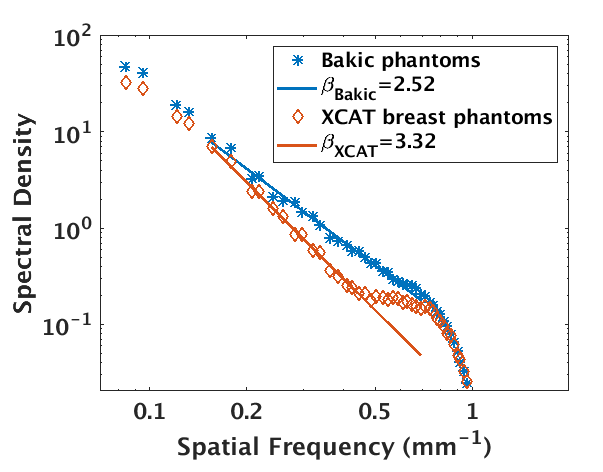} 
\caption{The power spectra analysis of DBT slices of both type of backgrounds for a sample acquisition protocol of $60^\circ$ arc span, 35 projections suggest that lack of small and sharp structures in XCAT breast phantoms resulted in lower spectral density at higher frequencies than that of Bakic phantoms.} 
\label{fig:nps_phtms}
\end{figure}

In order to characterize the structural variations of the selected phantom types, we estimated the NPS of the simulated DBT images. We selected the NPS of in-plane DBT images corresponding to the acquisition configurations of 35 projections over a $60^\circ$ arc span for comparing the phantoms' structures. Figure \ref{fig:nps_phtms} shows the averaged NPS of both the phantom types in log-log scale along with linear fits.
These plots show that Bakic phantom backgrounds have higher spectral densities at higher frequencies (considered as the anatomical region) than that of XCAT breast phantom backgrounds.
We note that this high anatomic noise is not reflected in the calculated $\beta$ values. The estimated values of $\beta$ were 2.52 and 3.32 respectively for Bakic and XCAT breast phantom backgrounds. 
This result is contradictory to popular belief that lower $\beta$ values indicate lower anatomical noise and in agreement with observations made in our prior work\cite{kavuri2020relative}.


\subsection{Gaze Analysis}

\begin{figure}[h!]
     \centering
	\includegraphics[width=0.23\textwidth]{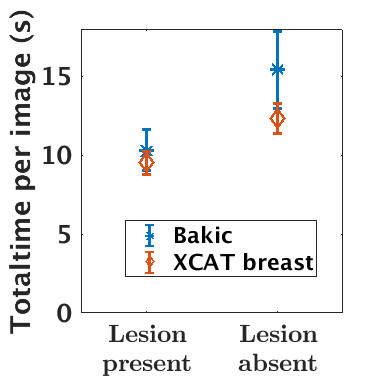}
	\includegraphics[width=0.23\textwidth]{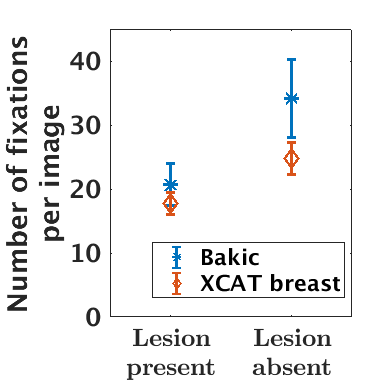}\vspace{\fill}
        \caption{\label{fig:Gze_obs}The average amount of time spent and the average number of fixations made on images (includes lesion absent and lesion present images) plotted for both phantom types. Observers spent longer time and made more fixations to make decisions on images with Bakic phantom backgrounds in comparison to those with XCAT breast backgrounds.}
\end{figure}

\begin{figure}[h!]
     \centering
	\includegraphics[width=0.45\textwidth]{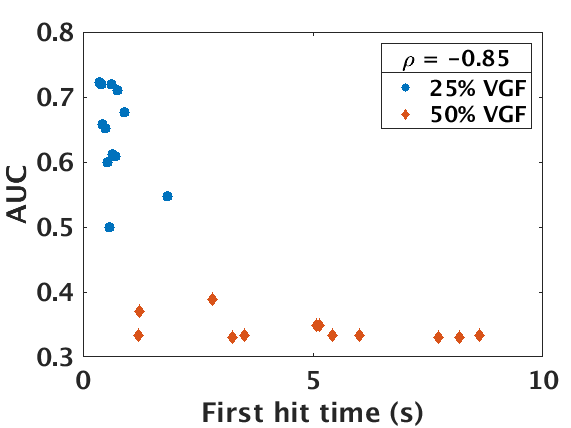}\vspace{\fill}\\
	\includegraphics[width=0.45\textwidth]{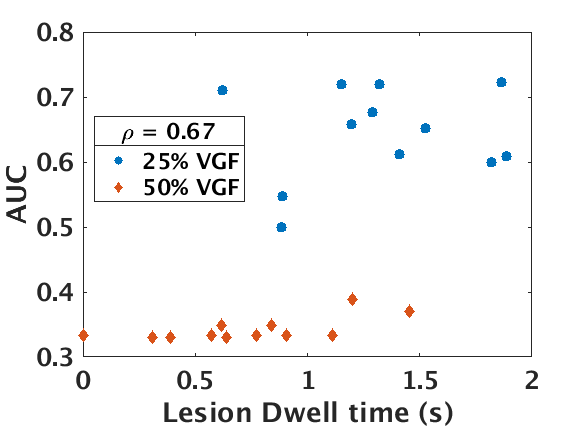}\vspace{\fill}\\
	\includegraphics[width=0.45\textwidth]{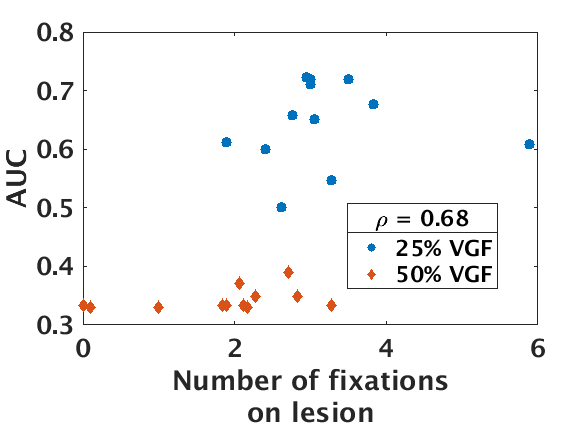}
        \caption{\label{fig:Gze_pht}Average value of first hit time, lesion dwell time and number of fixations on lesion were plotted against AUC value of each observer (total 6) for both phantom types and two densities.}
\end{figure}

 From the eye-tracking data, the total time spent as well as the total number of fixations made on images (both lesion present and lesion absent images) were estimated. For lesion present images, we also estimated the first hit time (on the lesion), lesion dwell time and the number of fixations on lesion. Each of these gaze metric was averaged across all images for a given phantom type and breast density for each observer. 
Figure \ref{fig:Gze_obs} shows the influence of phantom type on two of the gaze metrics plotted separately for lesion absent and lesion present images. Observers spent longer time and made more fixations to diagnose lesion absent images than lesion present images, which is in-line with previous findings \cite{timberg2013investigation,suwa2001analyzing}. The most striking result to emerge from this data is that observers spent longer time and made more fixations on images with the Bakic phantom backgrounds than XCAT breast backgrounds. This difference is significant ($p-value < 0.01$) in lesion absent images only.
A possible explanation for this difference may be that the greater anatomical noise of Bakic phantom backgrounds than the XCAT breast backgrounds (see Fig.\ref{fig:nps_phtms}) made observers less confident and hence spent longer time to make decision. 

Spearman rank correlation coefficient was computed between the average values of each gaze metric and the AUC values of both phantom types and densities. Table \ref{tab:cc_gaze} shows that all of the estimated gaze metrics have good correlation with diagnostic performance. 
Our observation of first hit time, lesion dwell time, and number of fixations on lesion showed that the observers took longer to first fixate on the lesion, spent less cumulative time on lesion, and had fewer fixations on the lesion as the task difficulty increased due to higher breast density. No significant difference was observed in these gaze metrics due to change of phantom type in lesion present images. This result indicates that lesion visibility or task difficulty is a major factor in defining the gaze pattern in lesion present images where as anatomical complexity is the major factor influencing the gaze pattern in lesion absent images. 
First hit time showed a strong negative (-0.85) correlation with the AUC values as shown in figure \ref{fig:Gze_pht}. This result suggest that the quicker an observer fixate on the lesion the better the diagnostic performance which is in accordance with the observation made by Kundel et al \cite{kundel2008using}. The positive correlation of lesion fixations and dwell time with AUC suggests that observers fixate the lesion longer and multiple times to locate accurately. All the five gaze metrics showed significant correlation ($p-value<0.01$) with AUC values.

\begin{table}[ht]
\centering
\caption{Spearman rank correlation between the AUC values of the six observers and the corresponding average values of the five gaze metrics. All five metrics show good and significant correlation with the diagnostic performance.}
\label{tab:cc_gaze}
\begin{tabular}{|c|c|c|}
\hline
\textbf{Gaze metric} & \textbf{Correlation coefficient ($\rho$)}  & \textbf{p-value} \\
 \hline
Total Number of Fixations & -0.52 & 0.0085 \\
Total time & -0.6 & 0.0019  \\
Lesion Dwell Time & 0.67 & 0.0003 \\
Lesion Number of fixations & 0.68 & 0.0003 \\
First Hit Time & \bf{ -0.85} & 1.2e-7 \\
\hline
\end{tabular}
\end{table}
\section{DISCUSSION AND CONCLUSION}
This work evaluated the influence of variations between digital breast phantoms on DBT optimization and interpretation process for a small lesion localization task. Our results indicate that 
the phantoms should have adequate structures at spatial frequencies that are relevant for the signal size and the intended task for sufficient realism.
We observed that optimal number of projections for peak detection performance could change with the structural complexity of the phantoms. 
In addition, we observed that the number of projections required to achieve maximum performance is smaller for XCAT breast phantoms than for Bakic phantoms. Our power spectrum analysis revealed that the complex structures in Bakic phantoms contribute to high frequencies while the high-frequency content of XCAT breast phantoms showed lower amplitudes. These high-frequency structures resulted in more aliasing artifacts in Bakic phantom backgrounds compared to XCAT breast phantom backgrounds under similar sparse sampling conditions. Hence, more projections were required to resolve these high-frequency structures in Bakic phantom backgrounds.

Our gaze analysis suggest that all the five gaze metrics showed good correlation with diagnostic performance, which is in accordance with the observations made by Voisin at al \cite{voisin2013investigating} in mammograpy. This result indicates that gaze metrics and interpretation process were influenced by the task difficulty. Hence, these gaze metrics can be used to understand the task difficulty in VIT imaging.

In this study, we evaluated the optimal number of projections in an arc span of $60^\circ$, which is wider than the arc ranges used in the clinical DBT systems. The arc samplings between the adjacent projections for peak performance were approximately $2.5^\circ$ and $6^\circ$ in Bakic and XCAT breast phantoms respectively for 3-mm lesion detection task. For other arc spans, we expect similar amount of aliasing artifacts for the similar arc sampling. Thus our conclusions should hold for other arc spans used in clinical DBT prototypes which have not been examined here in our VIT study. 

The difference in the optimal arc sampling from VIT conducted with two widely used phantom types, raises the need to standardize and unify the frequency contents and complexity of digital phantoms if results published from multiple groups need to be compared. The goal of our study was not to determine if one of the two phantoms are favorable or better than the other for use in VIT for breast. It is likely that the perceived realism and correlation with observer performance for VIT with actual clinical imaging trials would show results in favor of one of these phantoms based on the chosen task and the signal type. 

One of the limitations of our study is that only 3 observers participated in estimating AUC trends. This number is smaller than the number of observers participated in DBT optimization studies in literature\cite{zeng2017optimization,hadjipanteli2019threshold,goodsitt2014digital,vancoillie2021impact}.
However, the observers had higher agreement with each other which is backed by our estimation of intraclass correlation coefficient value of greater than 0.92. Secondly, we also estimated correlation between the average AUC trends of Bakic and XCAT breast phantoms. The spearman rank correlation of 0.2 suggests that the two trends are not similar. 

Furthermore, the study was conducted only for the task of searching and locating a 3-mm spherical lesion as conducting human observer studies for different signal sizes and types is time consuming and expensive. Prior studies suggest that smaller (high frequency) lesions dictate the optimal number of projections because their visibility is affected more by both aliasing artifacts (at lower number of projections) and random noise (mainly at higher number of projections) \cite{reiser2010task, gang2010anatomical}.
Many studies also used 3-mm \cite{sechopoulos2009optimization,gang2010anatomical,chawla2009optimized} and similar sizes \cite{zeng2017optimization,goodsitt2014digital,vancoillie2021impact,ikejimba2021assessment} of lesions as a signal target in optimization studies. 
We believe that for larger signals the differences in optimal arc sampling due to differences in the selected phantom structures may be less evident than that of smaller signals.
We chose a sphere lesion as target because detecting a sphere lesion is relatively easy task compared to complex spicule lesion study, hence requiring minimal training. In addition, the performance of non-radiologists and radiologists was shown to be similar for relatively easy tasks \cite{bertram2013effect} and multiple studies have used spherical lesions in virtual imaging trials\cite{sechopoulos2009optimization,gang2010anatomical,chawla2009optimized, reiser2010task,zeng2017optimization}. On the other hand, for high contrast smaller signals such as micro-calcification, random noise was shown to be the dominating factor in determining the optimal configuration rather than the background structures.\cite{reiser2010task} Future studies will include signals of different sizes and types.

Another limitation of our study is that only six phantoms (to generate a much larger number of independent cases in each study set) were selected for each phantom type, this sample may not represent the entire population of each phantom type. Although, our goal was not to compare between Bakic and XCAT phantoms, but to validate how optimization estimation changes with phantoms that were generated differently.  One may chose 6 phantoms in their studies in evaluating optimal configurations as some of the studies in literature chose around 2 to 9 phantoms \cite{gifford2016visual,das2008evaluation,hadjipanteli2019threshold,goodsitt2014digital,vancoillie2021impact}. 
If the selected new phantoms have similar NPS properties, we anticipate the similar differences in the optimal number of projections. Instead, if the selected phantoms have different NPS properties and results in different optimal number of projections, which strengthens our argument that phantom structures influence the estimation of system optimal configurations. 
The selected phantoms (25\% and 50\% dense) are in the higher range of breast densities observed in the clinical data.  The average AUC trends (see figure \ref{fig:auc_fig}) did not change due to change of breast density, suggesting that AUC trends may remain same for other density phantoms. These results also point to the need to include more phantoms with variations in VIT studies.

In conclusion, our results indicate that both the structural complexity and the relevant spatial frequency magnitudes in digital phantom structures can influence the estimation of optimal system configurations. Our results highlight the importance of accurate modeling of phantoms to resemble the patient anatomy and the importance of assessment for their realism (for chosen tasks) before use in VITs. This results can be generalized to digital phantoms used for 
multiple imaging modalities. As a final note, our goal in this particular study was not to discuss superiority of one phantom type over the other based on results shown here. The key aspect is for the VIT and medical physics community to be mindful that optimization or other results shown using one "realistic" breast phantom may not always agree with results when using another "realistic" phantom unless additional standardization efforts are pursued.

\section{ACKNOWLEDGEMENTS}
This work was partially supported by funding from the NIH National Institute of Biomedical Imaging and
Bioengineering (NIBIB) grant R01 EB EB029761, the US Department of Defense (DOD) Congressionally Directed Medical Research Program (CDMRP) Breakthrough Award BC151607 and the National Science Foundation CAREER Award 1652892. We would also like to thank both UPenn and Duke groups for making the digital phantoms freely available for research. 

\bibliography{references}

\begin{thebibliography}{10}

\bibitem{bakic2013mo}
P~Bakic, K~Myers, I~Reiser, N~Kiarashi, and R~Zeng.
\newblock Virtual tools for validation of x-ray breast imaging systems.
\newblock {\em Medical Physics}, 40(6Part23):390--390, 2013.

\bibitem{badano2017silico}
Aldo Badano, Andreu Badal, Stephen Glick, Christian~G Graff, Frank Samuelson,
  Diksha Sharma, and Rongping Zeng.
\newblock In silico imaging clinical trials for regulatory evaluation: initial
  considerations for victre, a demonstration study.
\newblock In {\em Medical Imaging 2017: Physics of Medical Imaging}, volume
  10132, page 1013220. International Society for Optics and Photonics, 2017.

\bibitem{das2011comparison}
Mini Das and Howard~C Gifford.
\newblock Comparison of model-observer and human-observer performance for
  breast tomosynthesis: effect of reconstruction and acquisition parameters.
\newblock In {\em Medical Imaging 2011: Physics of Medical Imaging}, volume
  7961, page 796118. International Society for Optics and Photonics, 2011.

\bibitem{das2015examining}
Mini Das, Zhihua Liang, and Howard~C Gifford.
\newblock Examining wide-arc digital breast tomosynthesis: optimization using a
  visual-search model observer.
\newblock In {\em SPIE Medical Imaging}, pages 94121S--94121S. International
  Society for Optics and Photonics, 2015.

\bibitem{barufaldi2018openvct}
Bruno Barufaldi, David Higginbotham, Predrag~R Bakic, and Andrew~DA Maidment.
\newblock Openvct: a gpu-accelerated virtual clinical trial pipeline for
  mammography and digital breast tomosynthesis.
\newblock In {\em Medical Imaging 2018: Physics of Medical Imaging}, volume
  10573, page 1057358. International Society for Optics and Photonics, 2018.

\bibitem{sharma2019silico}
Diksha Sharma, Christian~G Graff, Andreu Badal, Rongping Zeng, Purva Sawant,
  Aunnasha Sengupta, Eshan Dahal, and Aldo Badano.
\newblock In silico imaging tools from the victre clinical trial.
\newblock {\em Medical physics}, 46(9):3924--3928, 2019.

\bibitem{abadi2020virtual}
Ehsan Abadi, William~P Segars, Benjamin~MW Tsui, Paul~E Kinahan, Nick Bottenus,
  Alejandro~F Frangi, Andrew Maidment, Joseph Lo, and Ehsan Samei.
\newblock Virtual clinical trials in medical imaging: a review.
\newblock {\em Journal of Medical Imaging}, 7(4):042805, 2020.

\bibitem{barufaldi2021virtual}
Bruno Barufaldi, Andrew~DA Maidment, Magnus Dustler, Rebecca Axelsson, Hanna
  Tomic, Sophia Zackrisson, Anders Tingberg, and Predrag~R Bakic.
\newblock Virtual clinical trials in medical imaging system evaluation and
  optimisation.
\newblock {\em Radiation Protection Dosimetry}, 2021.

\bibitem{kavuri2020relative}
Amar Kavuri and Mini Das.
\newblock Relative contributions of anatomical and quantum noise in signal
  detection and perception of tomographic digital breast images.
\newblock {\em IEEE transactions on medical imaging}, 39(11):3321--3330, 2020.

\bibitem{kavuri2018interaction}
Amar Kavuri, Nathaniel~R Fredette, and Mini Das.
\newblock Interaction of anatomic and quantum noise in dbt power spectrum.
\newblock In {\em Medical Imaging 2018: Image Perception, Observer Performance,
  and Technology Assessment}, volume 10577, page 105770G. International Society
  for Optics and Photonics, 2018.

\bibitem{gifford2016visual}
Howard~C Gifford, Zhihua Liang, and Mini Das.
\newblock Visual-search observers for assessing tomographic x-ray image
  quality.
\newblock {\em Medical physics}, 43(3):1563--1575, 2016.

\bibitem{lau2013towards}
Beverly~A Lau, Mini Das, and Howard~C Gifford.
\newblock Towards visual-search model observers for mass detection in breast
  tomosynthesis.
\newblock In {\em Medical Imaging 2013: Physics of Medical Imaging}, volume
  8668, page 86680X. International Society for Optics and Photonics, 2013.

\bibitem{jiang2017analyzing}
Zhengqiang Jiang, Mini Das, and Howard~C Gifford.
\newblock Analyzing visual-search observers using eye-tracking data for digital
  breast tomosynthesis images.
\newblock {\em JOSA A}, 34(6):838--845, 2017.

\bibitem{gifford2008optimizing}
HC~Gifford, CS~Didier, Mini Das, and SJ~Glick.
\newblock Optimizing breast-tomosynthesis acquisition parameters with scanning
  model observers.
\newblock In {\em Medical Imaging 2008: Image Perception, Observer Performance,
  and Technology Assessment}, volume 6917, pages 227--235. SPIE, 2008.

\bibitem{das2008evaluation}
Mini Das, Howard Gifford, Michael O'Connor, and Stephen Glick.
\newblock Evaluation of a variable dose acquisition methodology for breast
  tomosynthesis.
\newblock In {\em Medical Imaging 2008: Physics of Medical Imaging}, volume
  6913, page 691319. International Society for Optics and Photonics, 2008.

\bibitem{das2009evaluation}
Mini Das, Howard~C Gifford, J~Michael o'connor, and Stephen~J Glick.
\newblock Evaluation of a variable dose acquisition technique for
  microcalcification and mass detection in digital breast tomosynthesis.
\newblock {\em Medical physics}, 36(6Part1):1976--1984, 2009.

\bibitem{gifford2018assessment}
Howard~C Gifford and Mini Das.
\newblock Assessment of dbt acquisition parameters for 2d and 3d search tasks.
\newblock In {\em Medical Imaging 2018: Image Perception, Observer Performance,
  and Technology Assessment}, volume 10577, page 105770I. International Society
  for Optics and Photonics, 2018.

\bibitem{nisbett2020correlation}
William~H Nisbett, Amar Kavuri, and Mini Das.
\newblock On the correlation between second order texture features and human
  observer detection performance in digital images.
\newblock {\em Scientific Reports}, 10(1):1--14, 2020.

\bibitem{nisbett2018investigating}
William~H Nisbett, Amar Kavuri, and Mini Das.
\newblock Investigating the contributions of anatomical variations and quantum
  noise to image texture in digital breast tomosynthesis.
\newblock In {\em Medical Imaging 2018: Physics of Medical Imaging}, volume
  10573, page 105730H. International Society for Optics and Photonics, 2018.

\bibitem{andrade2022sources}
Diego Andrade, Amar Kavuri, and Mini Das.
\newblock Sources of image texture variation in tomographic breast imaging.
\newblock In {\em Medical Imaging 2022: Image Perception, Observer Performance,
  and Technology Assessment}. SPIE, 2022.

\bibitem{glick2018advances}
Stephen~J Glick and Lynda~C Ikejimba.
\newblock Advances in digital and physical anthropomorphic breast phantoms for
  x-ray imaging.
\newblock {\em Medical physics}, 45(10):e870--e885, 2018.

\bibitem{sechopoulos2009optimization}
Ioannis Sechopoulos and Caterina Ghetti.
\newblock Optimization of the acquisition geometry in digital tomosynthesis of
  the breast.
\newblock {\em Medical physics}, 36(4):1199--1207, 2009.

\bibitem{gang2010anatomical}
GJ~Gang, DJ~Tward, J~Lee, and JH~Siewerdsen.
\newblock {Anatomical background and generalized detectability in tomosynthesis
  and cone-beam CT}.
\newblock {\em Medical physics}, 37(5):1948--1965, 2010.

\bibitem{lau2012statistically}
Beverly~A Lau, Ingrid Reiser, Robert~M Nishikawa, and Predrag~R Bakic.
\newblock A statistically defined anthropomorphic software breast phantom.
\newblock {\em Medical physics}, 39(6Part1):3375--3385, 2012.

\bibitem{baneva2017evaluation}
Yanka Baneva, Kristina Bliznakova, Lesley Cockmartin, Stoyko Marinov, Ivan
  Buliev, Giovanni Mettivier, Hilde Bosmans, Paolo Russo, Nicholas Marshall,
  and Zhivko Bliznakov.
\newblock Evaluation of a breast software model for 2d and 3d x-ray imaging
  studies of the breast.
\newblock {\em Physica Medica}, 41:78--86, 2017.

\bibitem{bakic2002mammogram}
Predrag~R Bakic, Michael Albert, Dragana Brzakovic, and Andrew~DA Maidment.
\newblock Mammogram synthesis using a 3d simulation. i. breast tissue model and
  image acquisition simulation.
\newblock {\em Medical physics}, 29(9):2131--2139, 2002.

\bibitem{bliznakova2010evaluation}
K~Bliznakova, Sankararaman Suryanarayanan, Andrew Karellas, and N~Pallikarakis.
\newblock Evaluation of an improved algorithm for producing realistic 3d breast
  software phantoms: Application for mammography.
\newblock {\em Medical physics}, 37(11):5604--5617, 2010.

\bibitem{chen2011anthropomorphic}
Baiyu Chen, Jamie Shorey, Robert~S Saunders~Jr, Samuel Richard, John Thompson,
  Loren~W Nolte, and Ehsan Samei.
\newblock An anthropomorphic breast model for breast imaging simulation and
  optimization.
\newblock {\em Academic radiology}, 18(5):536--546, 2011.

\bibitem{mahr2011three}
David~M Mahr, Rohit Bhargava, and Michael~F Insana.
\newblock Three-dimensional in silico breast phantoms for multimodal image
  simulations.
\newblock {\em IEEE transactions on medical imaging}, 31(3):689--697, 2011.

\bibitem{elangovan2017design}
Premkumar Elangovan, Alistair Mackenzie, David~R Dance, Kenneth~C Young,
  Victoria Cooke, Louise Wilkinson, Rosalind~M Given-Wilson, Matthew~G Wallis,
  and Kevin Wells.
\newblock Design and validation of realistic breast models for use in multiple
  alternative forced choice virtual clinical trials.
\newblock {\em Physics in Medicine \& Biology}, 62(7):2778, 2017.

\bibitem{carton2014virtual}
Ann-Katherine Carton, Anthony Grisey, Pablo Milioni~de Carvalho, Clarisse
  Dromain, and Serge Muller.
\newblock A virtual human breast phantom using surface meshes and geometric
  internal structures.
\newblock In {\em International Workshop on Digital Mammography}, pages
  356--363. Springer, 2014.

\bibitem{hsu2013generation}
Christina~ML Hsu, Mark~L Palmeri, W~Paul Segars, Alexander~I Veress, and
  James~T Dobbins~III.
\newblock Generation of a suite of 3d computer-generated breast phantoms from a
  limited set of human subject data.
\newblock {\em Medical physics}, 40(4):043703, 2013.

\bibitem{erickson2016population}
David~W Erickson, Jered~R Wells, Gregory~M Sturgeon, Ehsan Samei, James~T
  Dobbins~III, W~Paul Segars, and Joseph~Y Lo.
\newblock Population of 224 realistic human subject-based computational breast
  phantoms.
\newblock {\em Medical physics}, 43(1):23--32, 2016.

\bibitem{sturgeon2016eigenbreasts}
Gregory~M Sturgeon, Daniel~J Tward, M~Ketcha, JT~Ratnanather, MI~Miller, Subok
  Park, WP~Segars, and Joseph~Y Lo.
\newblock Eigenbreasts for statistical breast phantoms.
\newblock In {\em Medical Imaging 2016: Physics of Medical Imaging}, volume
  9783, pages 584--592. SPIE, 2016.

\bibitem{chen2017high}
Xinyuan Chen, Xiaolin Gong, Christian~G Graff, Maira Santana, Gregory~M
  Sturgeon, Thomas~J Sauer, Rongping Zeng, Stephen~J Glick, and Joseph~Y Lo.
\newblock High-resolution, anthropomorphic, computational breast phantom:
  fusion of rule-based structures with patient-based anatomy.
\newblock In {\em Medical Imaging 2017: Physics of Medical Imaging}, volume
  10132, pages 464--469. SPIE, 2017.

\bibitem{sarno2021dataset}
Antonio Sarno, Giovanni Mettivier, Francesca di~Franco, Antonio Varallo,
  Kristina Bliznakova, Andrew~M Hernandez, John~M Boone, and Paolo Russo.
\newblock Dataset of patient-derived digital breast phantoms for in silico
  studies in breast computed tomography, digital breast tomosynthesis, and
  digital mammography.
\newblock {\em Medical Physics}, 48(5):2682--2693, 2021.

\bibitem{chawla2009optimized}
Amarpreet~S Chawla, Joseph~Y Lo, Jay~A Baker, and Ehsan Samei.
\newblock Optimized image acquisition for breast tomosynthesis in projection
  and reconstruction space.
\newblock {\em Medical physics}, 36(11):4859--4869, 2009.

\bibitem{michael2013generation}
J~Michael~O'Connor, Mini Das, Clay~S Dider, Mufeed Mahd, and Stephen~J Glick.
\newblock Generation of voxelized breast phantoms from surgical mastectomy
  specimens.
\newblock {\em Medical physics}, 40(4):041915, 2013.

\bibitem{o2010development}
J~Michael O’Connor, Mini Das, Clay Didier, Mufeed Mah’d, and Stephen~J
  Glick.
\newblock Development of an ensemble of digital breast object models.
\newblock In {\em International Workshop on Digital Mammography}, pages 54--61.
  Springer, 2010.

\bibitem{bakic2011development}
Predrag~R Bakic, Cuiping Zhang, and Andrew~DA Maidment.
\newblock Development and characterization of an anthropomorphic breast
  software phantom based upon region-growing algorithm.
\newblock {\em Medical physics}, 38(6Part1):3165--3176, 2011.

\bibitem{cockmartin2013comparative}
Lesley Cockmartin, Hilde Bosmans, and NW~Marshall.
\newblock Comparative power law analysis of structured breast phantom and
  patient images in digital mammography and breast tomosynthesis.
\newblock {\em Medical physics}, 40(8):081920, 2013.

\bibitem{badano2017much}
Aldo Badano.
\newblock “how much realism is needed?”—the wrong question in silico
  imagers have been asking.
\newblock {\em Medical Physics}, 44(5):1607--1609, 2017.

\bibitem{reiser2010task}
I~Reiser and RM~Nishikawa.
\newblock Task-based assessment of breast tomosynthesis: Effect of acquisition
  parameters and quantum noise a.
\newblock {\em Medical physics}, 37(4):1591--1600, 2010.

\bibitem{zeng2017optimization}
Rongping Zeng, Aldo Badano, and Kyle~J Myers.
\newblock Optimization of digital breast tomosynthesis (dbt) acquisition
  parameters for human observers: effect of reconstruction algorithms.
\newblock {\em Physics in Medicine \& Biology}, 62(7):2598, 2017.

\bibitem{park2010statistical}
Subok Park, Robert Jennings, Haimo Liu, Aldo Badano, and Kyle Myers.
\newblock A statistical, task-based evaluation method for three-dimensional
  x-ray breast imaging systems using variable-background phantoms.
\newblock {\em Medical Physics}, 37(12):6253--6270, 2010.

\bibitem{zhao2014noise}
Z~Zhao, GJ~Gang, and JH~Siewerdsen.
\newblock Noise, sampling, and the number of projections in cone-beam ct with a
  flat-panel detector.
\newblock {\em Medical physics}, 41(6Part1):061909, 2014.

\bibitem{rajagopal2018evaluation}
Jayasai Rajagopal, Gregory~M Sturgeon, Xinyuan~C Chen, Thomas~J Sauer, Yinhao
  Ren, WP~Segars, and Joseph~Y Lo.
\newblock Evaluation of statistical breast phantoms with higher resolution.
\newblock In {\em Medical Imaging 2018: Physics of Medical Imaging}, volume
  10573, page 1057307. International Society for Optics and Photonics, 2018.

\bibitem{siewerdsen1997empirical}
JH~Siewerdsen, LE~Antonuk, Y~El-Mohri, J~Yorkston, W~Huang, JM~Boudry, and
  IA~Cunningham.
\newblock Empirical and theoretical investigation of the noise performance of
  indirect detection, active matrix flat-panel imagers (amfpis) for diagnostic
  radiology.
\newblock {\em Medical physics}, 24(1):71--89, 1997.

\bibitem{vedula2003computer}
Aruna~A Vedula, Stephen~J Glick, and Xing Gong.
\newblock Computer simulation of ct mammography using a flat-panel imager.
\newblock In {\em Medical Imaging 2003: Physics of Medical Imaging}, volume
  5030, pages 349--361. International Society for Optics and Photonics, 2003.

\bibitem{das2011penalized}
Mini Das, Howard~C Gifford, J~Michael O'Connor, and Stephen~J Glick.
\newblock Penalized maximum likelihood reconstruction for improved
  microcalcification detection in breast tomosynthesis.
\newblock {\em IEEE Transactions on Medical Imaging}, 30(4):904--914, 2011.

\bibitem{boone1999glandular}
John~M Boone.
\newblock Glandular breast dose for monoenergetic and high-energy x-ray beams:
  Monte carlo assessment.
\newblock {\em Radiology}, 213(1):23--37, 1999.

\bibitem{johns1987x}
Paul~C Johns and Martin~J Yaffe.
\newblock X-ray characterisation of normal and neoplastic breast tissues.
\newblock {\em Physics in Medicine \& Biology}, 32(6):675, 1987.

\bibitem{siddon1985fast}
Robert~L Siddon.
\newblock Fast calculation of the exact radiological path for a
  three-dimensional ct array.
\newblock {\em Medical physics}, 12(2):252--255, 1985.

\bibitem{lim1990two}
Jae~S Lim.
\newblock Two-dimensional signal and image processing.
\newblock {\em Englewood Cliffs, NJ, Prentice Hall, 1990, 710 p.}, 1990.

\bibitem{vieira2013effect}
Marcelo Andrade da~Costa Vieira, Predrag~R Bakic, Andrew Douglas~Arnold
  Maidment, et~al.
\newblock Effect of denoising on the quality of reconstructed images in digital
  breast tomosynthesis.
\newblock In {\em Proceedings of SPIE}. SPIE, 2013.

\bibitem{feldkamp1984practical}
Lee~A Feldkamp, LC~Davis, and James~W Kress.
\newblock Practical cone-beam algorithm.
\newblock {\em Josa a}, 1(6):612--619, 1984.

\bibitem{brunye2019review}
Tad~T Bruny{\'e}, Trafton Drew, Donald~L Weaver, and Joann~G Elmore.
\newblock A review of eye tracking for understanding and improving diagnostic
  interpretation.
\newblock {\em Cognitive research: principles and implications}, 4(1):1--16,
  2019.

\bibitem{kundel1978visual}
Harold~L Kundel, Calvin~F Nodine, and Dennis Carmody.
\newblock Visual scanning, pattern recognition and decision-making in pulmonary
  nodule detection.
\newblock {\em Investigative radiology}, 13(3):175--181, 1978.

\bibitem{kundel2008using}
Harold~L Kundel, Calvin~F Nodine, Elizabeth~A Krupinski, and Claudia
  Mello-Thoms.
\newblock Using gaze-tracking data and mixture distribution analysis to support
  a holistic model for the detection of cancers on mammograms.
\newblock {\em Academic radiology}, 15(7):881--886, 2008.

\bibitem{voisin2013investigating}
Sophie Voisin, Frank Pinto, Songhua Xu, Garnetta Morin-Ducote, Kathy Hudson,
  and Georgia~D Tourassi.
\newblock Investigating the association of eye gaze pattern and diagnostic
  error in mammography.
\newblock In {\em Medical Imaging 2013: Image Perception, Observer Performance,
  and Technology Assessment}, volume 8673, page 867302. International Society
  for Optics and Photonics, 2013.

\bibitem{olsen2012tobii}
Anneli Olsen.
\newblock The tobii i-vt fixation filter.
\newblock {\em Tobii Technology}, 21, 2012.

\bibitem{mackenzie2021effect}
Alistair Mackenzie, Sukhmanjit Kaur, Emma~L Thomson, Melissa Mitchell,
  Premkumar Elangovan, Lucy~M Warren, David~R Dance, and Kenneth~C Young.
\newblock Effect of glandularity on the detection of simulated cancers in
  planar, tomosynthesis, and synthetic 2d imaging of the breast using a hybrid
  virtual clinical trial.
\newblock {\em Medical Physics}, 48(11):6859--6868, 2021.

\bibitem{timberg2013investigation}
Pontus Timberg, Kristina L{\aa}ng, Marcus Nystr{\"o}m, Kenneth Holmqvist,
  Philippe Wagner, Daniel F{\"o}rnvik, Anders Tingberg, and Sophia Zackrisson.
\newblock Investigation of viewing procedures for interpretation of breast
  tomosynthesis image volumes: a detection-task study with eye tracking.
\newblock {\em European radiology}, 23(4):997--1005, 2013.

\bibitem{suwa2001analyzing}
Koji Suwa, Akira Furukawa, Toru Matsumoto, and Takashi Yosue.
\newblock Analyzing the eye movement of dentists during their reading of ct
  images.
\newblock {\em Odontology}, 89(1):0054--0061, 2001.

\bibitem{hadjipanteli2019threshold}
Andria Hadjipanteli, Premkumar Elangovan, Alistair Mackenzie, Kevin Wells,
  David~R Dance, and Kenneth~C Young.
\newblock The threshold detectable mass diameter for 2d-mammography and digital
  breast tomosynthesis.
\newblock {\em Physica Medica}, 57:25--32, 2019.

\bibitem{goodsitt2014digital}
Mitchell~M Goodsitt, Heang-Ping Chan, Andrea Schmitz, Scott Zelakiewicz,
  Santosh Telang, Lubomir Hadjiiski, Kuanwong Watcharotone, Mark~A Helvie,
  Chintana Paramagul, Colleen Neal, et~al.
\newblock Digital breast tomosynthesis: studies of the effects of acquisition
  geometry on contrast-to-noise ratio and observer preference of low-contrast
  objects in breast phantom images.
\newblock {\em Physics in Medicine \& Biology}, 59(19):5883, 2014.

\bibitem{vancoillie2021impact}
Liesbeth Vancoillie, Lesley Cockmartin, Nicholas Marshall, and Hilde Bosmans.
\newblock The impact on lesion detection via a multi-vendor study: A
  phantom-based comparison of digital mammography, digital breast
  tomosynthesis, and synthetic mammography.
\newblock {\em Medical Physics}, 48(10):6270--6292, 2021.

\bibitem{ikejimba2021assessment}
Lynda~C Ikejimba, Jesse Salad, Christian~G Graff, Mitchell Goodsitt, Heang-Ping
  Chan, Hailiang Huang, Wei Zhao, Bahaa Ghammraoui, Joseph~Y Lo, and Stephen~J
  Glick.
\newblock Assessment of task-based performance from five clinical dbt systems
  using an anthropomorphic breast phantom.
\newblock {\em Medical physics}, 48(3):1026--1038, 2021.

\bibitem{bertram2013effect}
Raymond Bertram, Laura Helle, Johanna~K Kaakinen, and Erkki Svedstr{\"o}m.
\newblock The effect of expertise on eye movement behaviour in medical image
  perception.
\newblock {\em PloS one}, 8(6):e66169, 2013.

\end{thebibliography}
\bibliographystyle{unsrt}

\end{document}